\documentclass{article}
\usepackage{spconf,amsmath,graphicx}
\usepackage{tablefootnote}
\usepackage{xcolor}

\title{Analyzing the robustness of unsupervised speech recognition}
\name{Guan-Ting Lin$^{1 *}$\thanks{$^*$ These authors contributed equally.}, Chan-Jan Hsu$^{1,2 *}$, Da-Rong Liu$^{1 }$, Hung-Yi Lee$^{1 }$, Yu Tsao$^{2 }$}
\address{$^{1 }$National Taiwan University, Taiwan  \\
$^{2 }$Academia Sinica, Taiwan \\
\{r10942104, r09946011, hungyilee\}@ntu.edu.tw\thanks{This material is based on work that is partially funded by an unrestricted gift from Google.}}

\begin{document}
\ninept
\maketitle
\begin{abstract}
Unsupervised speech recognition (unsupervised ASR) aims to learn the ASR system with non-parallel speech and text corpus only. Wav2vec-U \cite{baevski2021unsupervised} has shown promising results in unsupervised ASR by self-supervised speech representations coupled with Generative Adversarial Network (GAN) training, but the robustness of the unsupervised ASR framework is unknown. In this work, we further analyze the training robustness of unsupervised ASR on the domain mismatch scenarios in which the domains of unpaired speech and text are different. Three domain mismatch scenarios include: (1) using speech and text from different datasets, (2) utilizing noisy/spontaneous speech, and (3) adjusting the amount of speech and text data. We also quantify the degree of the domain mismatch by calculating the JS-divergence of phoneme n-gram between the transcription of speech and text. This metric correlates with the performance highly. Experimental results show that domain mismatch leads to inferior performance, but a self-supervised model pre-trained on the targeted speech domain can extract better representation to alleviate the performance drop.
\end{abstract}
\section{Introduction}
\label{sec:intro}
Automatic speech recognition (ASR) is a long-standing research area that predicts transcriptions given the input speech feature. The performance of ASR has rapidly improved due to the deep learning method with an increasing amount of labeled data. However, collecting hundreds or thousands of hours of paired speech and text data is costly and even infeasible for the low-resource or endangered languages. 

On the other hand, unsupervised ASR aims only to leverage unpaired speech and text data for training an ASR system. Unsupervised ASR systems learn the cross-modal mapping between speech and text \cite{yeh2018unsupervised, aldarmaki2021unsupervised, Liu2020TowardsUS,liu2018completely,chen2019completely}. 
In particular, \cite{liu2018completely} and \cite{chen2019completely} propose to learn the mapping by Generative Adversarial Network (GAN) \cite{goodfellow2014generative}. They demonstrate encouraging results for completely unsupervised ASR training, but the performance is still far from the supervised ASR system. Notably, a recent work named Wav2vec-U \cite{baevski2021unsupervised}  has shown remarkable performance breakthroughs in unsupervised ASR. The performance is comparable with some supervised methods \cite{amodei2016deep, zhang2020pushing, xu2018neural}.

\begin{figure}[h]
\includegraphics[width=8cm]{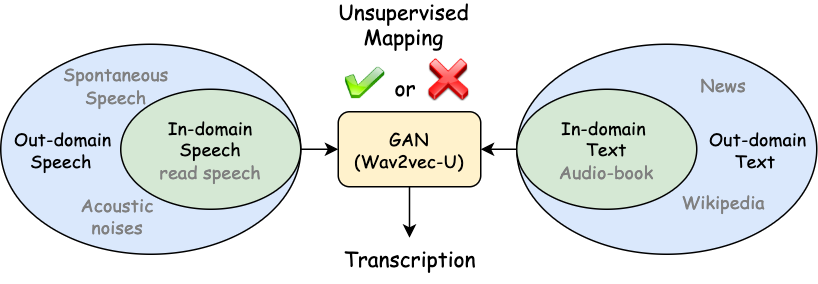}
\vspace{-0.2cm}
\caption{A brief illustration of our work.}
\vspace{-0.3cm}
\label{fig:main}
\end{figure}
This paper intends to investigate the robustness of Wav2vec-U under three different domain mismatch scenarios, where the domain of unpaired speech and text are different. Figure \ref{fig:main} shows the main concept of our work. The three scenarios are as below:\\
\textbf{Content dissimilarity}: 
One major experiment in Wav2vec-U is trained with an unpaired speech from Librispeech \cite{7178964} and text from Librispeech language modeling corpus, which both come from the audiobook but are collected in speech and text form, respectively. The experimental setup is unrealistic in the unsupervised learning scenario because no prior knowledge about speech transcription exists. The intrinsic properties of the selected text corpus and speech data may be reasonably different.
Therefore, we investigate the robustness of Wav2vec-U by utilizing the speech and text that come from different datasets to simulate dissimilarities in content. \\
\textbf{Spontaneous speech}:
The audio data used in the training of Wav2vec-U, even in low-resource experiments, are all read speech.
Read speech is clean and well-structured;
on the other hand, real-world speech data contains noises and disfluencies. Furthermore, real-world speech lacks literary expressions and causes dissimilarity to text corpus in content.  
To investigate the robustness of Wav2vec-U with real-world speech, we conduct further experiments on spontaneous speech datasets such as conversational speech and live talk, which is real-time generated with the casual styles. 
\\
\textbf{Quantity limitation}: 
The results of Wav2vec-U show that it is possible to train an unsupervised ASR model by a minimal amount of speech and text from the audiobook. However, in the domain-mismatched conditions, whether we can still use a limited amount of data to train an unsupervised ASR model is under-explored. The more significant variance present in fewer data intensifies the degree of
domain mismatch. Hence, we restrict the amount of data in domain mismatch conditions for testing the training robustness. 

\begin{table*}[h]
    \centering
    \begin{tabular}{|c|c|c|c|c|}
        \hline
        Corpus&LibriLM&Wiki&NewsCrawl&ImageC \\
        \hline
        Domain & Literature&Encyclopedia&News&Image Caption \\
        \hline
        Size (word) &635M&82M&432M&33M  \\
        \hline
        Description & Librispeech language  &  Extracted from&Extracted article text & Combination of COCO \cite{lin2014microsoft}\\
         &  modeling data with the   & the set of verified &from multiple online news & and Conceptual Caption \cite{sharma2018conceptual}   \\ 
          & Libri-Light removed \cite{synnaeve2019end}$^{1}$ & ``Good" and ``Featured"   &publications$^{2}$ & containing annotated and\\ 
           & &articles on Wikipedia \cite{wiki} & & web-scraped image captions. \\
           \hline
    \end{tabular}
    \vspace{-0.2cm}
    \caption{Descriptions of text corpora.}
    \label{tab:text}
\end{table*}

\begin{table*}[h]
  \centering
    \begin{tabular}{ |c|c|c|c|c| }
    \hline
    Name & Domain & Size (hr) & Speech Characteristics \\
    \hline
    Librispeech train-960 \cite{7178964} & Read Literature & 960 &  Read speech, speak clearly \\ 
    SwitchBoard \cite{switchboard} & Telephone conversation & 300 &  Spontaneous, noisy, speak fast \\
    TED-LIUM v3 \cite{Hernandez_2018} & Live talk & 452 & Spontaneous, little noise,  speak clearly\\
    \hline
    \end{tabular}
    \vspace{-0.2cm}
    \caption{Descriptions of speech dataset}
    \label{tab:speech}
\end{table*}

The main findings of our work are listed below: 
\begin{itemize}
    \item When the content of speech and text is dissimilar, using a large amount of text retains the performance, but reducing to the small amount of texts considerably degrades the results.
    \item We discover that 4-gram information distribution of phoneme is critical for successfully training the GAN-based unsupervised ASR model.
    \item Self-supervised pre-training on targeted speech domain enables extracting better feature representation and significantly improves the performance.
\end{itemize}
\label{sec:format}
\section{GAN-based Unsupervised ASR}
Unsupervised ASR takes audio signals as input and outputs the phoneme sequences without phoneme-labeled supervision.  
\cite{liu2018completely} first proposes to cluster the speech embedding into discrete tokens and applies Generative Adversarial Network \cite{goodfellow2014generative} to learn a mapping matrix that aligns discrete acoustic sequences and phoneme sequences.  Although this pioneering work shows encouraging preliminary results, the method requires annotations of phoneme boundaries. \cite{chen2019completely} propose to use unsupervised audio segmentation approaches as the initial segmentation, then adopting Hidden Markov Model (HMM)/GAN harmonization to refine the segmentation boundaries iteratively. This method does not need any annotated phoneme boundaries and can even improve performance. 

Recently, Wav2vec-U \cite{baevski2021unsupervised} has shown promising results in unsupervised speech recognition using self-supervised speech representation coupled with GAN training and self-training. Specifically, they extract acoustic representation by the self-supervised pre-trained model, Wav2vec 2.0 \cite{w2v2}, which consists of a convolutional feature encoder to convert raw waveform into latent representation and a Transformer \cite{vaswani2017attention} for learning context representation by the self-supervised objective. Next, they cluster the representations extracted by Wav2vec 2.0, apply PCA to reduce dimension, and mean-pool the representations in the same cluster to generate condensed segment representations. They apply GAN to map the segment representation to phoneme prediction, and the preliminary phoneme prediction has already outperformed previous work by a large margin. Finally, to further refine the prediction, they adopt self-training for HMM, which retrains HMMs by treating GAN outputs as pseudo labels or fine-tuning another Wav2vec 2.0 model by pseudo labels. After subsequent self-training, the error rate is close to the supervised ASR model. \renewcommand\thefootnote{\textcolor{white}{\arabic{footnote}}}
\footnote{$^{1}$https://github.com/flashlight/wav2letter/tree/main/recipes/sota/2019} 
\footnote{$^{2}$http://www.statmt.org/wmt14/translation-task.html}
\renewcommand\thefootnote{\textcolor{black}{\arabic{footnote}}}

\vspace{-0.8cm}
\section{Experimental setup}
\begin{table*}[t]
  \small
  \centering
  
    \begin{tabular}{ |c|c|c|c|c|c|c|c| }
    \hline
    \multicolumn{2}{|c|}{Speech} & \multicolumn{6}{|c|}{Text} \\
    \hline
    Corpus & Hour & LibriLM & Wiki & NewsCrawl & ImageC & matched* & unmatched* \\
    \hline
    \multicolumn{8}{|c|}{\textit{Full amount of speech}} \\
    \hline
    Librispeech train & 960 & 20.25 &26.02 &21.83 &31.59 & N/A  & N/A  \\
    TED-LIUM v3 & 452 & 31.62 & 35.21  & 32.05  & 41.87
  & 28.13  & N/A \\
    SwitchBoard  & 300 & 92.10  & 93.08  & 95.25  & 80.15 & 35.80  & N/A  \\
    SwitchBoard-\textit{w2v2-all}  & 300 & 44.38  & 94.12 & 43.44  & 72.10  & 32.34  & N/A  \\
    
    \hline
    \multicolumn{8}{|c|}{\textit{Little amount of speech}} \\
    \hline
    
    Librispeech train & 9.6 &22.51 &29.03&24.65&105.00 &- &- \\
    TED-LIUM v3 & 10 & 36.01&88.26 
&33.52 
&85.92 &29.13 &32.44 \\
    SwitchBoard  & 10 &  95.86 & - & - & - & 95.13 & 93.48 \\
    SwitchBoard-\textit{w2v2-all}  & 10 &  92.10 & - & - & -& 96.14 & 93.48\\

    \hline
    \end{tabular}
    \vspace{-0.2cm}
    \caption{Phoneme Error Rate (PER) of different speech-text corpus pairs. If not specified, we use \textit{w2v2} for feature extraction. We use the entire text data of the text corpus in Table \ref{tab:text}. A dashed line denotes that we do not experiment on that combination because experimental results in the most feasible setting (speech/in-domain exact matched text) have already failed to converge (higher than 90\% PER). Therefore, we expect those speech and out-of-domain text combinations to fail with a PER around 90, and the exact numerical PER result is no longer meaningful in that range. N/A denotes that the combination is non-applicable.}
    \label{tab:PER}
\end{table*}
\begin{table*}[t]
 
  \centering
    \begin{tabular}{
    |c|c|c|c|c|c|c|c| }
    
    \hline
    \multicolumn{2}{|c|}{Speech} & \multicolumn{6}{|c|}{Text} \\
    \hline
    Name & Hour & LibriLM & Wiki & NewsCrawl & ImageC & matched* & unmatched* \\
    \hline
    Librispeech train & 960 &  0.0058&0.1697&0.0785&0.2635& N/A & N/A
    
\\ 
    Librispeech train & 9.6 &0.0720&0.2228&0.1375&0.3003&-&-\\
    SwitchBoard & 300&  0.1357
&0.2513&0.1499
&0.3174
&0&N/A \\
    SwitchBoard & 10&  0.2248&0.3339&0.2407&0.3869&0&0.0866 \\
    TED-LIUM v3 & 452 & 0.0880&0.1825&0.0841& 0.2678&0& N/A\\
    TED-LIUM v3 & 10 & 0.1909&0.2855&0.1928&0.3452&0& 0.1014 \\
    \hline
    \end{tabular}
    \vspace{-0.2cm}
    \caption{4-gram JS-Divergence of different speech-text corpus pairs. We use the entire text data of the text corpus in Table \ref{tab:text}. N/A denotes that the combination is non-applicable}
    \vspace{-0.4cm}
    \label{tab:JSD} 
\end{table*}

The primary interest of this work is to examine the robustness of the state-of-the-art unsupervised ASR method, Wav2vec-U. Our work focuses on how domain mismatch affects cross-modal GAN training, so we do not use tricks like HMM self-training and fine-tuning for refinement. We measure the phoneme error rate (PER) of the Viterbi-decoded generator output.

\subsection{Data preparation}
We designed three main testing setups, and the data information is in Table \ref{tab:text} and \ref{tab:speech}. \\
\textbf{Content dissimilarity}:
We test the GAN performance when the text corpus represents different domains, including encyclopedia (Wiki), News (NewsCrawl), and image captions (ImageC). The vocabulary usages and narrative styles are pretty dissimilar among them. For instance, the text in News is used to report current affairs, whereas the ImageC text consists of descriptions of images. The sentence length for ImageC is short, and adjectives and nouns consist of a larger proportion. \\
\textbf{Spontaneous speech}:
We regard spontaneous speech as the domain mismatched speech because it is naturally generated and relatively easy to collect. We choose the SwitchBoard dataset to represent telephone/conversational speech and the TED-LIUM v3 dataset for lectures/monologue. The quality of the TED talk speech is clean and includes multiple types of oratory skills by the speakers. In contrast, the speech in the SwitchBoard contains telephone channel noise, disfluency, and various speaking styles. \\
\textbf{Quantity limitation}:
\cite{baevski2021unsupervised} successfully trained Wav2vec-U with 9.6 hours of LibriSpeech and 3k text sentences in LibriLM. We are interested in how the low amount of speech and text data affects performance when the content of speech and text are mismatched. Therefore, we cap the speech data to 10 hours and the text data to 3k or 30k sentences and repeat experiments in the first two settings.
\subsection{Implementation Details}
We train unsupervised speech recognition using the official code\footnote{https://github.com/pytorch/fairseq/tree/main/examples/wav2vec\\/unsupervised}. 
All the English text corpora are transformed to phoneme by grapheme-to-phoneme conversion \cite{g2pE2019} and pre-processed with 0.25 rate of silence token insertion as \cite{baevski2021unsupervised}. The speech pre-processing involves silence removal by the unsupervised robust voice activity detection (rVAD)  \cite{rVAD} and audio representation extracted by self-supervised pre-training speech model. Specifically, we leverage two kinds of pre-trained Wav2vec 2.0 Large model \cite{w2v2}. One is only pre-train on reading speech from Libri-Light \cite{librilight} (\textit{\textbf{w2v2}}), the other is pre-train on heterogeneous sources, including Libri-Light, CommonVoice \cite{ardila-etal-2020-common}, Switchboard \cite{switchboard} and Fisher \cite{fisher1}\cite{fisher2}) from \cite{hsu2021robust} (\textit{\textbf{w2v2-all}})\footnote{Both pre-trained models' checkpoint are open-source and can be downloaded on  https://github.com/pytorch/fairseq/tree/main/examples/wav2vec}. 

The training and evaluation are based on the standard training and validation split for Librispeech and TED-LIUM v3. There is no standard split for Switchboard, so we create the training/validation set by 80/20 random sampling ratio. The content of two settings: ``matched" and ``unmatched" depends on the speech dataset; the matched set is the transcription of the speech dataset, while the unmatched setting is the transcription of the unused portion in the speech dataset. If the entire audio dataset is used for training, such as the Librispeech train-960 dataset, the ``unmatched" set does not exist.
For hyperparameter settings, we set code penalty to 4, gradient penalty to 2, and smoothness weight to 0.5. We train 150000 steps with a batch size of 128. Following the unsupervised model selection metric in Wav2vec-U, we use the perplexity of the phoneme language model to choose the best model. Each setup is trained with three random seeds and reports the average PER.
\subsection{Domain mismatch measurement}
To quantify the content difference between speech and text, we use the phoneme sequence of speech transcription and text to count the n-grams, then converting counting into n-gram phoneme distributions. Finally, we measure the n-gram Jensen-Shannon Divergence (JSD)  between the two distributions to estimate the extent of domain mismatch between speech and text content. 

JSD is a smoothed symmetric version of the Kullback–Leibler divergence (KLD), and the formula for JSD between two distributions P and Q is:
\[JSD(P||Q)=\frac{1}{2}  KLD(P||\frac{Q+P}{2})+\frac{1}{2} KLD(Q||\frac{Q+P}{2})\] with
\[KLD(P||Q) = \sum{P(x)log\frac{P(x)}{Q(x)}}\]

\section{Experimental Results}
We conclude the experimental results with three critical questions about Wav2vec-U training robustness.
\vspace{-0.5cm}
\subsection{How do different text influence the performance?}
We apply JSD to measure the degree of content dissimilarity. Table \ref{tab:JSD} shows the 4-gram JSD of experimented speech-text corpora pairs. For the matched setup, the JSD is always zero. As expected, the JSD between Librispeech and LibriLM is very low. For spontaneous speech such as TED-LIUM and SwitchBoard, pairing with both LibriLM and NewsCrawl produce similar low scores. The 4-gram distribution is larger when the text corpus is the Wiki or ImageC, suggesting their content is more dissimilar to spontaneous speech. 

\begin{figure}[t]
\includegraphics[width=7cm]{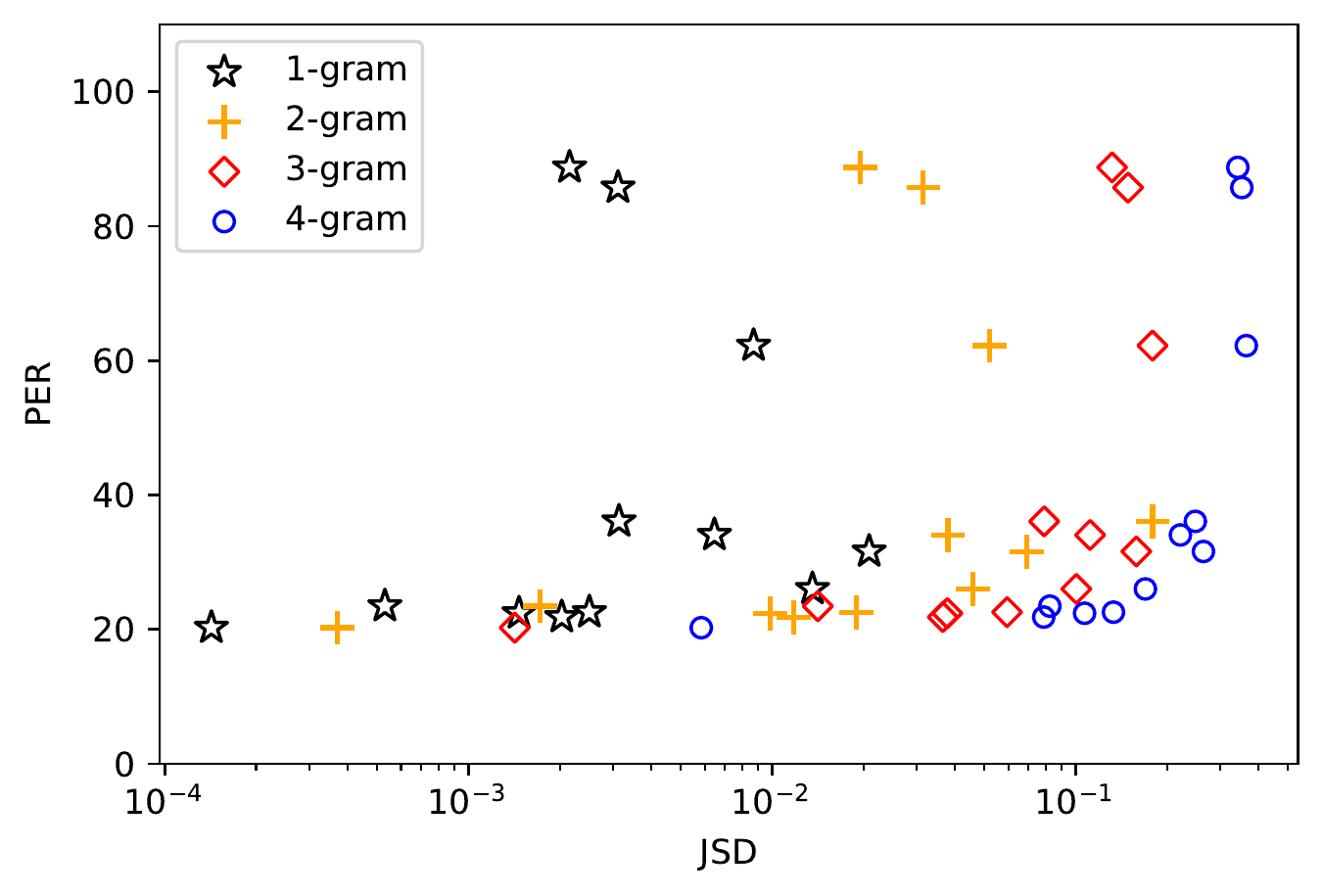}
\centering
\vspace{-0.2cm}
\caption{JSD-PER between Librispeech-960 and text.  Each set of 1 to 4-gram connected by one horizontal PER line represents one experimental setting. For all experiments, the corresponding text corpus can be found in Table \ref{tab:PER} and \ref{tab:text3000} using PER as the index.}
\label{fig:JSD}
\end{figure}

The results of PER under different settings are in Table \ref{tab:PER}.
Pairing the PER results in Table \ref{tab:PER} the corresponding JSD in Table \ref{tab:JSD}, we observe a correlation between n-gram JSD and the performance, so we plotted the n-gram JSD to PER for Librispeech-960 experiments in Figure \ref{fig:JSD}, where we analyze phoneme collocations from length 1 to 4. Our experiments observe that the PER rises sharply when 4-gram JSD increases to a certain degree. 
For 3-gram, the boundary is not perfect but provides rough guidance, while such a boundary is nonexistent in both 1-gram and 2-gram. 
We can therefore infer that the GAN training mainly relies on 3-gram and 4-gram information. 
Since the 4-gram JSD is the most discriminant among our experimented n-grams, we only focus on 4-grams in the following sections, and we refer  \textit{\textbf{trainability threshold}} as the 4-gram JSD value that corresponds to the large PER gap around 40 to 60. 
\subsection{Does spontaneous speech influence the performance?}
In the upper part of Table \ref{tab:PER} labeled with ``Full amount of speech", the resulting PER shows that we can successfully train GAN with TED talk speech paired with all kinds of text corpora. As for the results of Switchboard, when the feature representations are extracted from \textit{w2v2}, only training with matched text is successful. The undesirable performance may come from telephone noise, disfluency in spoken conversation, and diversity of speaking styles, which degrade the feature representation extracted from \textit{w2v2}. The poor representations result in less accurate speech segmentation and further hinder GAN training.
\begin{figure}[t]
\centering
\includegraphics[width=7.3cm]{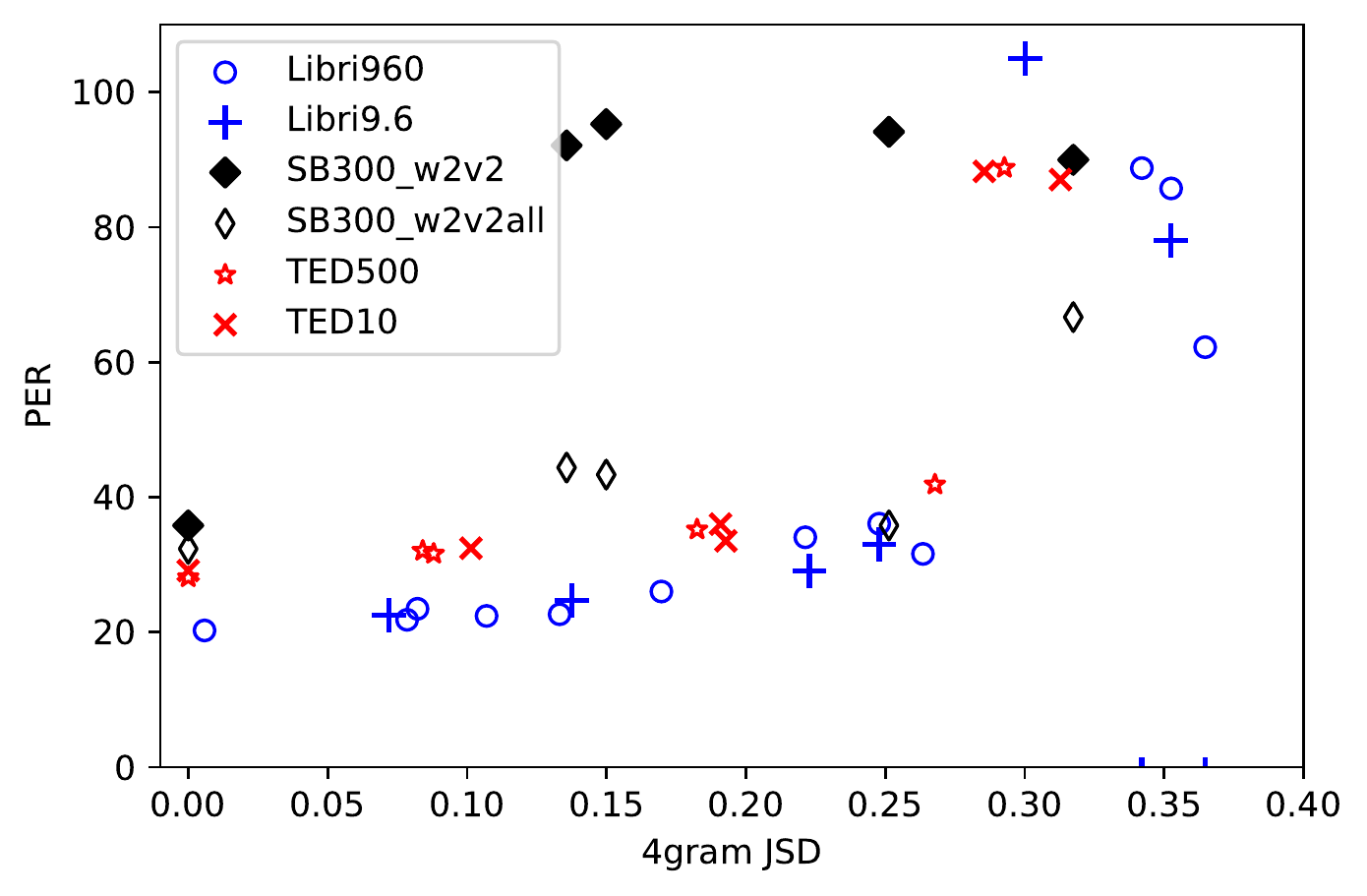}
\vspace{-0.2cm}
\caption{4-gram JSD-PER between Speech and text. Each notation corresponds to a kind of speech, while different points of the same notation denote different experimental settings. The detailed speech-text pairings can be found in Table \ref{tab:PER} and \ref{tab:text3000} using PER as the index.}
\label{fig:JSD-all}
\end{figure}
\begin{table}[t]
    \centering
    \begin{tabular}{|c|c|c|c|c|c|}
         \hline
         Text & Hour & LibriLM & Wiki & NewsCrawl & ImageC \\
         \hline
         30k & 960 & 23.46 & 22.56 &22.40&34.05   \\
         3k & 960 & 25.85 & 70.08 & 69.20 & 62.22 \\
         3k & 9.6 & 33.03 & - & - & - \\
         \hline
    \end{tabular}
    \vspace{-0.2cm}
    \caption{Phoneme Error Rate (PER) for Librispeech and the limited amount of text, only 3k or 30k sentences. A dashed line in the Table denotes that we do not experiment on that combination because the same scenarios with more speech training data have already failed.}
    \label{tab:text3000}
\end{table}
\begin{table}[t]
    \centering
    \begin{tabular}{|c|c|c|c|c|c|}
         \hline
         Text & Hour & LibriLM & Wiki & NewsCrawl & ImageC \\
         \hline
         30k & 960 & 0.0822 &  0.1332&0.1070&0.2213\\
         3k & 960 &  0.2478&0.3525&0.3420&0.3647 \\
         3k & 9.6 & 0.2607 &  0.3583&0.3492&0.3716 \\
         \hline
    \end{tabular}
    \vspace{-0.2cm}
    \caption{JSD for Librispeech and the limited amount of text, only 3k or 30k sentences for training.}
    \label{tab:text3000JSD}
\end{table}

We leverage \textit{w2v2-all} for feature extraction to extract better feature representations, which is pre-trained on broad domains, including Switchboard. Table \ref{tab:PER} demonstrates encouraging results that \textit{w2v2-all} improve PER from 92.10 to 44.38 with LibirLM and 95.23 to 43.44 with NewsCrawl. This result indicates that the quality of representation is crucial for training unsupervised ASR and self-supervised pre-training on target-domain speech helps extract better representations, significantly improving GAN performance. 

In the 4-gram JSD-PER plot of different speech data (Figure 2), the trainability threshold changes depending on the speech quality and extracted representation. For the Librispeech and TED talk speech data, the trainability threshold lies around 0.27.
On the other hand, for Switchboard data extracted by \textit{w2v2}, only training with matched text results in reasonable PER, so the trainability threshold is close to 0. When we extract features by \textit{w2v2-all}, the trainability threshold increases to around 0.25. We conclude that since TED-LIUM is clean spontaneous speech, the content similarity is the primary factor, while for noisy speech such as SwitchBoard, a better representation is required.
\subsection{Does the amount of data matter when the domain is mismatched?}
In the bottom part of Table \ref{tab:PER} labeled with ``Limited amount of speech", training with the limited amount of speech data produces inferior results to full-speech data setups, especially on the Switchboard dataset, where training fails across all text settings, including training with the ``matched" corpus with no divergence at all. Table \ref{tab:text3000JSD} shows that sub-sampling speech data increases JS-divergence and content dissimilarity given the same text corpus. 

The results of using the limited amount of text data are listed in Table \ref{tab:text3000} and \ref{tab:text3000JSD}. We can observe that training with full Librispeech (960 hours of speech data) and 3k text in domains other than LibriLM fails. The poor performance can be explained by a significant increase in JS-divergence value when the text dataset is limited to 3k sentences. When JSD exceeds its trainability threshold, the performance drops sharply as a result. Overall, the experimental results demonstrate that when speech and text data are already domain-mismatched, the amount of data plays an essential role in GAN training.
\section{Conclusion}
In this paper, we comprehensively investigate the robustness of the Wav2vec-U in different domain mismatch scenarios.
The experimental results indicate that content dissimilarity indeed degrades the performance, which correlates with the 4-gram JSD metric.
Besides, spontaneous speech makes the GAN training more difficult, but pre-training the self-supervised model on the target speech data dramatically reduces the phoneme error rate. In addition, when non-parallel speech and text are from different domains, more data is required to make unsupervised ASR feasible. We conclude that unsupervised ASR requires improvement in domain mismatch scenarios and hope to incentivize future research on this kind of direction.
\section{Acknowledgement}
We thank National Center for High-performance Computing (NCHC) of National Applied Research
Laboratories (NARLabs) in Taiwan for providing computational and storage resources.

\bibliographystyle{IEEEbib}
\bibliography{refs}

\end{document}